\documentstyle[12pt,twoside,fleqn,espcrc1,epsfig]{article}

\def\beq{\begin{equation}}
\def\eeq{\end{equation}}
\def\ba{\begin{eqnarray}}
\def\ea{\end{eqnarray}}

\def\v8p{v_8^\prime}

\newcommand{\AmS}{{\protect\the\textfont2
  A\kern-.1667em\lower.5ex\hbox{M}\kern-.125emS}}

\title{Neutron Matter : A Superfluid Gas}

\author{S. -Y. Chang $^a$, J. Morales, Jr. $^a$, V. R. Pandharipande $^a$, D. G. Ravenhall 
\address{ Department of Physics,  University of Illinois at Urbana-Champaign, 
         Urbana, IL 61801}, J. Carlson \address{ Theoretical Division, 
	 Los Alamos National Laboratory, Los Alamos, NM 87545},
         Steven C. Pieper $^c$, R. B. Wiringa \address{ Physics Division, Argonne 
         National Laboratory, Argonne, IL 60439} and K. E. Schmidt \address{
         Department of Physics and Astronomy, Arizona State University, 
         Tempe, AZ 85287}}

\begin{document}
% typeset front matter
\maketitle

\begin{abstract}
We review recent progress in the theory of neutron matter with particular 
emphasis on its superfluid properties.  Results of quantum Monte Carlo 
calculations of simple and realistic models of uniform superfluid neutron gas are 
discussed along with those of neutrons interacting in a potential well 
chosen to approximate neutron-rich oxygen isotopes. 
The properties of dilute superfluid Fermi gases that may be produced in atom 
traps, and their relations with neutron matter, are illustrated. 
The density dependence of the effective interaction between neutrons, 
used to describe neutron-rich systems with the mean field approximations, 
is also discussed.

\end{abstract}

\section{INTRODUCTION}

Neutron matter is believed to be an 
unusual superfluid gas with positive pressure at all 
densities.  It has been studied in the context of neutron stars for many 
years.  Recent reviews can be found in references \cite{anr1,anr2,rmp1}. 
However, the subject is rapidly progressing; the large number of authors 
in this paper represents the breadth of recent interest.  

The interesting properties of neutron matter follow from the fact that 
the $nn$ scattering length ($a=-18$ fm) has a large magnitude, so that 
the dimensionless parameter $|ak_F|\gg 1$ at densities   
$ > 10^{-4}\rho_0$, where $k_F$ is the Fermi momentum 
given by $k_F^3=3 \pi^2 \rho$.  We use $\rho_0=0.16$ fm$^{-3}$ to denote the 
density of nuclear matter in large nuclei.  On the other hand, the effective 
range of the $nn$ interaction ($r_{nn} \sim 2.8$ fm) is not much smaller than 
the interparticle distance $r_0 = (3 \rho/4 \pi)^{1/3}$, 
in neutron matter at densities of interest.  
Hence the $nn$ interaction can not be easily approximated by a zero range 
$\delta$-function interaction.  Realistic models of the $nn$ 
interaction have a repulsive core and strong tensor component; thus 
simple perturbation theory expansions can not be easily used to describe 
neutron matter.  In addition, due to the strong $nn$ attraction in $^1S_0$ 
channel, the low-density neutron matter is believed to be an $S-$wave 
superfluid, although at higher densities it could be a $P-$wave superfluid 
\cite{rmp1}.  

Recently it has become possible to study cold dilute Fermi gases in atom traps 
\cite{atomt}.  In these gases the interatomic interaction can be approximated 
by a zero range potential characterized by the scattering length $a$. 
The experiments have the ability to tune values of $ak_F$ by using Feshbach 
resonances.  These studies essentially probe superfluid gases like neutron matter
at very low densities ($r_0 \gg 2.8$ fm). 

Density-dependent effective interactions are commonly used in nuclear physics 
to develop mean-field approximations based on energy-density functionals. 
In absence of experimental data, results of many-body calculations of the 
equation of state of pure neutron 
matter \cite{fp,wff} are often used to determine the effective interactions 
\cite{pr,sly4,faya}. 

In section 2 we briefly describe the progress in quantum Monte Carlo 
calculations of the properties of dilute Fermi gases and neutron matter. 
Section 3 is devoted to quantum Monte Carlo calculations of neutrons in a 
potential well aimed to approximate neutron-rich oxygen isotopes.  We 
discuss effective interactions in neutron matter in section 4, and 
conclude in section 5. 

\section{UNIFORM SUPERFLUID GASES AND NEUTRON MATTER}

Recently it has become possible to calculate the energies and gaps of simple 
superfluid Fermi gases {\em ab initio} from the interparticle interaction, 
using quantum Monte Carlo methods \cite{sfgprl}.  The total energy is 
calculated for $N$ particles in a periodic box of volume $N/\rho $.  The 
results for dilute (effective range $\ll r_0$) Fermi gas with $ak_F = \infty$ 
are shown in Fig.1.  The unit of energy:
\beq
E_{FG}=\frac{3}{10} \frac{\hbar^2}{m} k_F^2~,
\eeq
corresponds to the energy per particle of noninteracting Fermi gas.  Dimensional 
arguments show that the energy and gap of dilute superfluid gases scale with 
$E_{FG}$ when $a \rightarrow \infty$.  The pairing gap $\Delta$ is obtained from 
the difference between results for odd and even $N$ (Fig.1). 

\begin{figure}[t!]
\centerline
{\ \epsfig{figure=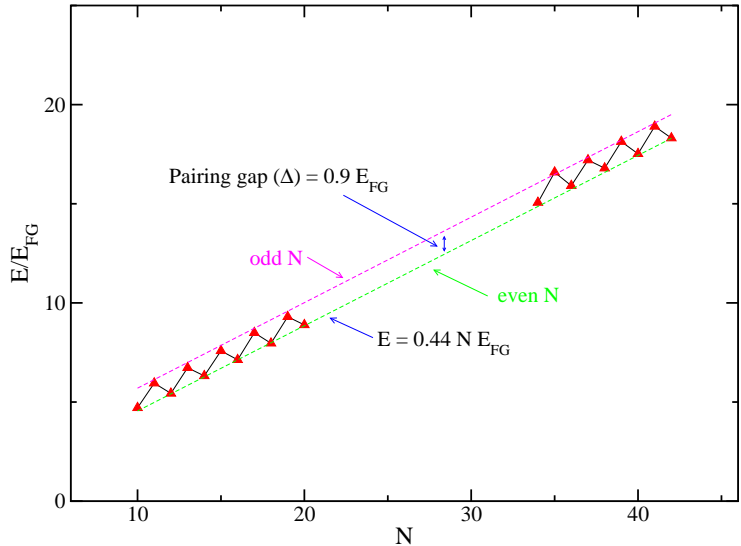,height=4.0in,angle=270}}
\caption[fig]{ GFMC results for the $E(N)$ of dilute superfluid gas with 
$ak_F \rightarrow \infty$, in a periodic box.}
\label{fig:sfge}
\end{figure}

These studies of dilute superfluid Fermi gases have been extended to various 
values of $ak_F$ in the domain $- \infty < (ak_F)^{-1} < \infty$.  Cubic periodic 
boxes have shell structure with shell closures at 2, 14, 38, ... for
spin-$\frac{1}{2}$ fermions.  In dilute gases the pairing removes shell 
effects when the interaction is attractive and strong enough so that 
$(ak_F)^{-1} > -$0.5.  The gap energy 
defined as:
\beq
\Delta(N)=E(N)-\frac{1}{2}[E(N-1)+E(N+1)]~,~~odd~~N~,
\eeq
does not depend significantly on $N$ for $(ak_F)^{-1} \geq -$1.  At negative values 
of $ak_F$, the calculated values of $\Delta$ are in between the estimates 
obtained from BCS and Gorkov equations \cite{sfgprl}.  
However, at large positive values of 
$(ak_F)^{-1}$, the gap energy is below the Gorkov estimate and approaches 
$|E_2|/2$, where $E_2$ is the energy of the two-body ground state.  Recall that when 
the interaction is attractive and $a > 0$, there must be one or more two-body 
bound states. 

As $(ak_F)^{-1}$ varies from $-\infty$ to $+\infty$ the character of dilute Fermi 
gases changes from BCS superfluids with weak pairing to diatomic Bose gases with 
Bose-Einstein condensation.  This transition has been discussed in the literature 
\cite{Leggett,Randeria}. 

The pairing in nuclei and neutron matter is much weaker than in the dilute 
Fermi gases with $(ak_F)^{-1} > -$0.3.  However, the energy of neutron matter 
appears to be close to the energy $(\sim 0.44 E_{FG})$ predicted by quantum 
Monte Carlo calculations (Fig.1) for large $|ak_F|$,  
over a wide density range. 
Most calculations of neutron matter with realistic interactions, carried out since 
1970 \cite{anr1}, give $E(\rho) \sim 0.5 E_{FG}(\rho)$ in the density range 
$0.01$ to $0.5 \rho_0$.  The pairing gaps in neutron matter at $\rho > 0.1 \rho_0$ 
are sensitive to the details of the $nn$ interaction.  For example, the gaps 
predicted from only the scattering length and effective range are much larger 
than those with realistic models with repulsive cores \cite{rmp1}. 

Quantum Monte Carlo calculations of neutron matter with realistic interactions 
are much more difficult than those for dilute gases.  In order to treat the effects of 
the spin, tensor and spin-orbit components in the $nn$ interaction 
exactly, one has to sum over all the $2^N$ spin states of the $N$ interacting 
neutrons \cite{nmgfmc}.  Such Green's function Monte Carlo (GFMC) calculations 
are limited to systems with no more than 16 neutrons at present.  Methods
to sample the spin space using auxiliary fields are also being developed. 
These auxiliary field diffusion Monte Carlo (AFDMC) calculations \cite{afdmc} 
%**** seem to be fairly accurate, and can be carried out for larger values of $N$. 
are fairly accurate for pure neutron systems, and can be carried out for much larger values of $N$. 

In computations using periodic boxes the interparticle interaction beyond
$L/2$, where $L$ is the box length, has to be truncated or approximated.  
GFMC calculations \cite{nmgfmc} use truncated interactions and estimate the 
contribution of $v(r > L/2)$ by approximate methods.  It is small for $\rho \leq 
\rho_0/2$, but increases rapidly with density.  The studies of pairing gaps 
in neutron matter have therefore focused on $\rho \leq \rho_0/4$.  The present,  
preliminary, results are listed in Table I. 

The effect of shell closure at $N=14$ is obvious in the results shown in 
Table I.  The energy of the seventh pair, $E(14)-E(12) \sim 3.2$ MeV, is much 
smaller than that of the eighth pair, $E(16)-E(14) \sim 26.8$ MeV.  Calculations 
with smaller statistical (Monte Carlo) error are necessary to study the dependence 
of $\Delta(N)$ on $N$, in addition to studies of $\Delta(N > 15)$ with the AFDMC 
method.  Nevertheless the $\Delta(\rho_0/4)$ appears to be 
of order $2.8(5)$ MeV.  This value is in excellent agreement with that obtained 
using realistic bare $nn$ interactions and the BCS equation, 2.6(1) MeV. 
This agreement is surprising since most estimates \cite{rmp1} suggested that 
medium polarization corrections to the bare interaction would suppress the 
gap by a factor of two or more even at a density of $\rho_0/4$. 

The BCS equation predicts a maximum gap of 3 MeV at $0.1\rho_0$, and the 
disappearance of the $^1S_0$ pairing at $\sim 0.7\rho_0$ \cite{rmp1}.  
It would be most 
interesting to verify these predictions with essentially exact {\em ab initio} 
calculations. 

\begin{table}
\caption{Preliminary results of quantum Monte Carlo calculations of the 
energies of $N$ neutrons at density $\rho_0/4$, in a cubic periodic box 
using the Argonne $v'_8$ interaction truncated at $L/2$, in MeV.}  
\vspace{0.2cm}
\begin{tabular}{ccccc}
   & GFMC &  GFMC &  AFDMC & AFDMC \\
$N$ & $E(N)$ & $\Delta(N)$ & $E(N)$ & $\Delta(N)$ \\ 
\hline
10  & 82.8(2) &    &  &   \\
11  & 86.8(2) &~~2.65(32) &  & \\
12  & 85.5(3) &    & 89.4  & \\
13  & 89.7(4) & 2.6(5)  &  94.0(4) & 2.6(5) \\
14  & 88.7(4) &   &  93.4(4)  &   \\
15  &     &     &~~110.1(3)~~&~~3.3(4)~~\\
16  &     &     &  120.2(2)  &   \\
\hline
\end{tabular}
%\label{table:params}
\end{table}

\section{Neutron Drops : Models of Neutron-Rich Nuclei}

Neutron drops bound in external wells were studied previously with various 
objectives \cite{ndrop1,ndrop2,ndrop3}.  We are presently exploring the 
possibility to model neutron-rich oxygen isotopes by neutron drops in a 
well meant to represent the effect of interactions with the protons.  The 
model Hamiltonian for the isotope with $N$ neutrons has the form:
\beq
H=\sum_{i=1,N}\left(-\frac{\hbar^2}{2m}\nabla^2_i+V(r_i)\right) 
  +\sum_{i<j\leq N} v_{ij} + \sum_{i<j<k \leq N} V_{ijk}~,
\eeq
where $v_{ij}$ and $V_{ijk}$ are realistic bare two-, and three-neutron 
interactions.  The Argonne $v_{18}$ and Illinois-2 interactions are used 
in the present calculations and the single-neutron potential $V(r)$ is 
chosen so that the density distribution of the eight-neutron ground state 
is close to that of protons in $^{16}$O.  It is given by:
\beq
V(r)=\frac{-35.5}{1+\rm{exp}[(r-3)/1.1]}~,
\eeq
in MeV and fm units.  This well has only two bound states; a 0s and a 0p 
at $\sim$ -12.1 and -4.8 MeV respectively.  It can thus bind only 8 noninteracting 
neutrons.  However, bound states of many more interacting neutrons exist in 
this well due to the $nn$ attraction. 

The calculated energies of the low-energy bound states of 
up to 14 neutrons are compared with the experimental energies of oxygen 
isotopes in Fig.2.  A constant of -19 MeV 
has been added to the calculated energies 
to match the energy of the 8-neutron drop with that of $^{16}$O. 
The results indicate that neutron-rich ($N > 8$) oxygen isotopes 
can be well approximated by neutron drops in a Woods-Saxon well.  
However, the approximation becomes less accurate for neutron-poor 
($N < 8$) isotopes.  Presumably the proton distribution in oxygen 
isotopes does not change much with $N$ when $N>8$, but when $N$ becomes 
smaller than 8, the protons become less bound and their distribution 
changes significantly. 

In the traditional shell model, neutron-rich oxygen isotopes are approximated 
by $N-8$ neutrons interacting via an effective interaction in the sd-shell. 
In contrast the present method uses bare interactions among all the neutrons.  
The single-particle 
potential $V(r)$ is meant to represent only the $np$ interaction 
effects.  It can obviously be improved by adding a spin-orbit term to the 
$V(r)$. 

In $^{17}$O the interactions between neutrons give only $\sim 2/3$ of the 
observed 
$d_{3/2}-d_{5/2}$ spin-orbit splitting, and the three-neutron interaction 
contributes approximately half of the calculated splitting.  The remaining third must come from the 
interactions with protons.  The variation of the pairing gap with $N$ is 
being studied in more detail. 

\begin{figure}[t!]
\centerline
{\ \epsfig{figure=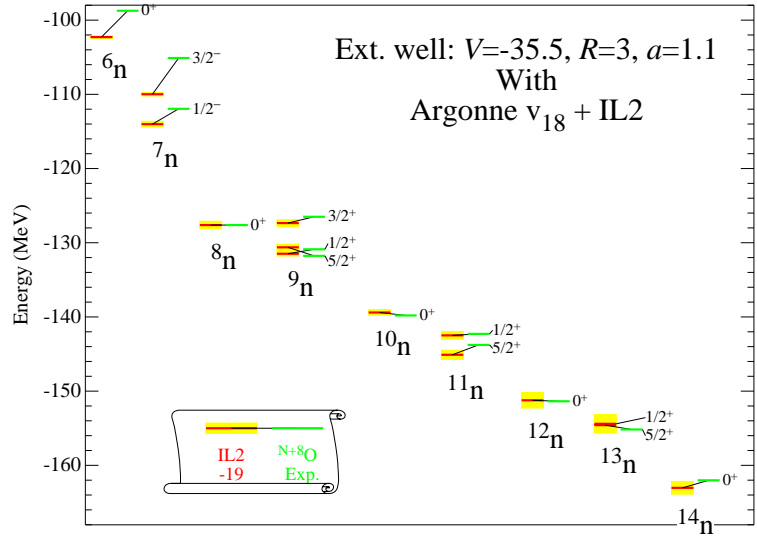,height=4.0in,angle=270}}
\caption[fig]{ The energies of neutron drops in the chosen external well are 
compared with the experimental energies of oxygen isotopes with the 
same number of neutrons.}
\label{fig:oxydr}
\end{figure}

\section{The Effective Interaction in Neutron Matter}

Density-dependent effective interactions are often used to describe nuclei 
using the mean-field approximation.  They are mostly used in the Hartree, 
Hartree-Fock or Hartree-Fock-Bogolubov equations.  In principle, parts of 
such effective interactions which depend only on the density of matter, can 
be chosen such as to reproduce the $E(\rho)$ and the pairing gap $\Delta(\rho)$ 
of uniform neutron matter.  

The simplest Skyrme effective interaction between neutrons is of the form:
\beq
v_{eff}({\bf R,r}) = b(\rho({\bf R}))~ \delta({\bf r})~.
\eeq
Here ${\bf R}$ and ${\bf r}$ are the center of mass and relative separation 
between the two interacting neutrons.  Commonly used Skyrme interactions have 
momentum dependent and spin-orbit terms which are omitted here for brevity.  Their 
dependence on the proton density can also be neglected in the present 
discussion.  In the 
Hartree-Fock approximation, the energy of neutron matter per neutron 
given by this $v_{eff}$ is:
\beq
E(\rho)=E_{FG}(\rho)+ \frac{1}{4} b(\rho) \rho~.
\eeq
The strength of the Skyrme interaction $b(\rho)$ can thus be obtained from the 
$E(\rho)$ of neutron matter. 

Most realistic calculations of neutron matter give $E(\rho) \sim 0.5 E_{FG}(\rho)$ 
in the density range $\rho=0.01$ to 0.1 fm$^{-3}$.  This result is independent of 
the interaction model and the calculation details \cite{anr1}.  It simply follows 
from the fact that in this density range the $|a_{nn}k_F| \gg 1$.  This implies that:
\beq
b(\rho) \sim - \frac{2}{\rho} E_{FG}(\rho) \sim \alpha \rho^{-1/3}~, 
\eeq
since $E_{FG}$ is proportional to $\rho^{2/3}$.  This low-density divergence of 
the effective $nn$ interaction is also a consequence of the large magnitude of the 
$nn$ scattering length. 

In some models \cite{pr} the neutron-proton effective interaction has a more severe 
$(1/\rho)$ divergence due to the existence of the $np$ bound state, the deuteron. 
However, this divergence 
is mostly ignored because matter containing neutrons and protons is 
unstable at low densities; it clusters into nuclei.  In contrast the $\rho^{-1/3}$ divergence 
of the effective $nn$ interaction could have consequences on the structure of 
neutron-rich nuclei with a neutron halo and in the neutron gas in a neutron-star crust 
\cite{anr1}.  

It is known that pairing is very important in determining the structure of neutron-rich 
nuclei.  For example, nuclei with an even number of neutrons dominate the neutron-drip 
line.  It is not possible to use delta-function effective interactions of finite 
strength to 
calculate pairing gaps without restricting the Hilbert space.  In unrestricted 
space divergences occur.  In stable nuclei reasonable values of the gaps may be 
obtained using the Skyrme effective interaction 
and truncating the Hilbert space to a single open shell \cite{ndrop2}. 
However, in nuclei near the neutron drip line, shell structure 
becomes less dominant, and the problem becomes more difficult.  Alternative 
approaches of using a local density approximation for the pairing fields 
\cite{Bulgac} are also being pursued. 

\section{Conclusions}

Advances in quantum Monte Carlo Methods in the new millennium have made possible essentially 
exact, {\em ab initio} calculations of the equations of state of low-density neutron 
matter.  We are also developing methods to calculate the pairing gaps in 
low-density neutron matter, and in dilute Fermi gases from the bare interactions.  
All realistic models of nuclear forces predict similar properties of low-density 
neutron matter, thus these {\em ab initio} predictions are expected to have little 
model dependence.  Ways to define effective interactions in neutron matter from the 
results of these {\em ab initio} calculations are being refined.  They can presumably 
be used to study the structure of neutron-rich nuclei.  In addition, studies of the 
bound states of neutrons in external potential wells meant to represent the effects 
of the neutron-proton interaction may also provide 
new insights in the structure of neutron-rich nuclei.

\section{ACKNOWLEDGMENTS}

The calculations were made possible by grants of time
on the parallel computers of Argonne and Los Alamos National Laboratory and 
the National Energy Research Scientific Computing Center.  The work of SYC,
JM, VRP, and DGR is supported by the  U.~S.~National Science Foundation
via grant PHY 00-98353, that of JC by
the U.~S.~Department of Energy under contract W-7405-ENG-36, and that
of SCP and RBW by the U.~S.~Department of Energy, Nuclear Physics
Division, under contract W-31-109-ENG-38.

\end{document}